\documentclass[aps,superscriptaddress,preprint]{revtex4}

\usepackage{amsmath,amssymb}
\usepackage{graphicx}
\usepackage{textcomp}
\usepackage{color}

\newcommand{\eq}[1]{Eq.~(\ref{#1})}
\newcommand{\fig}[1]{Fig.~\ref{#1}}

\renewcommand{\sec}[1]{Sec.~\ref{#1}}

\begin{document}

\title{Mesoscopic model for filament orientation in growing actin networks: the role of obstacle geometry}
\author{Julian Weichsel}
\email{weichsel@berkeley.edu}
\affiliation{Bioquant and Institute for Theoretical Physics, University of Heidelberg, Germany}
\affiliation{Department of Chemistry, University of California at Berkeley, United States}
\author{Ulrich S. Schwarz}
\email{ulrich.schwarz@bioquant.uni-heidelberg.de}
\affiliation{Bioquant and Institute for Theoretical Physics, University of Heidelberg, Germany}

\date{\today}

\begin{abstract}
Propulsion by growing actin networks is a universal mechanism used in
many different biological systems, ranging from the sheet-like
lamellipodium of crawling animal cells to the actin comet tails induced by certain bacteria and
viruses in order to move within their host cells.  Although the core
molecular machinery for actin network growth is well preserved in all
of these cases, the geometry of the propelled obstacle varies
considerably. During recent years, filament orientation distribution
has emerged as an important observable characterizing the structure and dynamical state of
the growing network. Here we derive several continuum equations for
the orientation distribution of filaments growing behind stiff
obstacles of various shapes and validate the predicted steady state
orientation patterns by stochastic computer simulations based on
discrete filaments. We use an ordinary differential equation approach
to demonstrate that for flat obstacles of finite size, two
fundamentally different orientation patterns peaked at either $\pm35$
or $+70/0/\!\!-\!\!70$ degrees exhibit mutually exclusive stability,
in agreement with earlier results for flat obstacles of very large lateral
extension. We calculate and validate phase diagrams as a function of
model parameters and show how this approach can be extended to
obstacles with piecewise straight contours. For curved obstacles, we arrive at a partial differential equation in the continuum limit,
which again is in good agreement with the computer simulations. In all
cases, we can identify the same two fundamentally different
orientation patterns, but only within an appropriate reference frame,
which is adjusted to the local orientation of the obstacle
contour. Our results suggest that two fundamentally different network
architectures compete with each other in growing actin networks,
irrespective of obstacle geometry, and clarify how simulated and
electron tomography data have to be analyzed for non-flat obstacle
geometries.
\end{abstract}

\maketitle



\section{Introduction}
\label{introduction}

The growth of actin networks is a generic propulsion mechanism
occurring in a large variety of biological systems, ranging from the
protruding lamellipodia of animal cells to the actin comet tails
recruited by pathogens like the bacterium {\it Listeria  monocytogenes} or the virus
{\it Vaccinia} within their host cells. Due to its central importance, the
molecular basis of this process is well preserved over a wide range of
different species \cite{carlier2010}. Although a large number of
accessory proteins is known to be involved in actin dynamics on the
cellular scale, in regions close to the leading edge the dynamics of
network growth are determined by a small number of key reactions. Here
the interplay between three fundamental processes determines the
structure of the growing network: polymerization of filamentous actin,
branching mediated by the protein complex Arp2/3 and binding of
capping proteins to the filament tip preventing further growth
\cite{mullins1998,pollard2007}. The fact that these processes are
highly conserved is impressively demonstrated by the observation that many
intracellular pathogens rely on them for efficient infection and
spread in the cytoplasm of their hosts
\cite{frischknecht2001,gouin2005,lambrechts2008}. Even functionalized
plastic beads, rods and discs as well as lipid vesicles and oil
droplet in purified protein solutions containing this minimal set of
molecules have been shown to be propelled \textit{in vitro} by polymerizing
actin networks
\cite{loisel1999,cameron1999,upadhyaya2003,giardini2003,schwartz2004,boukellal2004}.

While the underlying molecular basis is very similar in all of these
cases, the geometry of the different propelled objects is very
different. For instance, the highly curved shape of relatively small,
almost ellipsoidal pathogens like {\it Listeria  monocytogenes} or {\it Vaccinia}
  virus is very different from the relatively flat shape of the
leading edge of a crawling cells. The exact shape of the membrane in
migrating cells is very dynamic, but certainly corrugated on different
scales, thus in this case a flat obstacle shape can only be a first
approximation and corrugated obstacle contours are also of large
interest. In a growing actin network, new filaments nucleate by
branching off from existing mother filaments at a characteristic angle
around $70^\circ$ set by the molecular geometry of the Arp2/3-complex.
Importantly, branching can occur only close to the surface, as it
depends on the presence of surface-bound nucleation promoting factors
(NPF) like WASP or ActA \cite{beltzner2008}. Whether the daughter
filament becomes a productive member of the growing actin gel depends
on the way its direction is oriented relative to the direction of
growth and how the object is shaped.  Therefore obstacle geometry is a
crucial determinant of the resulting structural organization of the
network.

One essential function of growing actin networks is to generate force
against external load. Force-velocity relations for growing actin
networks have been measured in different experimental setups and for
various obstacle shapes
\cite{wiesner2003,mcgrath2003,marcy2004,parekh2005,prass2006,heinemann2011,zimmermann2012}. Unexpected
discrepancies in the results indicate that obstacle shape and the
resulting difference in network organization also has an effect on
force generation. Modern electron tomography leads to
visualization and analysis of filamentous actin networks in ever
greater detail \cite{resch2010,vinzenz2012}. While early electron
microscopy data for the lamellipodia of fish keratocytes suggested a
dendritic network of relatively short actin filaments
\cite{svitkina1999}, recent electron tomography data has revealed a
more diverse structural organization, with relatively long filaments
connected by few branch points into different and spatially extended
filament subsets \cite{urban2010,vinzenz2012}. In addition, it has
been demonstrated that the structural organization of the network
strongly depends on the protrusion speed, with fast growing networks
dominated by two symmetric diagonal filament orientations and slowly growing networks featuring
more filaments in parallel and orthogonal to the leading edge
\cite{koestler2008,weichsel2012}. It is to be expected, that in the
future the effect of obstacle geometry on network structure can be
quantified by such methods as well.

In order to achieve a complete understanding of the structure of
growing actin networks, different modeling approaches have been
developed during the last decade \cite{carlsson2010_2}, ranging
from microscopic ratchet models \cite{mogilner1996,mogilner2003}, rate
equations for filament growth
\cite{carlsson2003,maly2001,weichsel2010_2} and large-scale
computer simulations of filament ensembles
\cite{schaus2008,lee2009,alberts2004,schreiber2010} to continuum theories of how
elastic stress propels the obstacle \cite{gerbal2000} and multi-scale
models combining several of these model classes \cite{zimmermann2012,zhu2012}. From some of
these studies it has emerged that one central quantity characterizing
the structural organization of growing actin networks is the filament
orientation distribution, which can be directly compared with
experimental results
\cite{maly2001,verkhovsky2003,schaub2007,weichsel2012}.

In the following, we study a model which allows us to predict the
filament orientation distribution as a function of obstacle geometry.
Our reference point will be stochastic computer simulations of growing
actin networks incorporating the molecular processes of branching,
capping and filament polymerization \cite{weichsel2010_2}. For
computational simplicity and deeper insight, these will be compared to
different versions of a rate equation model
\cite{maly2001,carlsson2003,weichsel2010_2}. For actin growth behind
flat and laterally widely spread obstacles the steady state
organization has been predicted earlier to be either a $\pm35$ or
$+70/0/\!\!-\!\!70$ degree filament orientation pattern, with mutually
exclusive stability determined by the model parameters
\cite{maly2001,weichsel2010_2}.  In this paper, we will extend this
analysis to also predict the effect of finite size and geometry of the
propelled object. In particular, we will derive and validate a
partial differential equation, which is valid also for curved obstacle
shapes.

The article is organized as follows: In \sec{modeling_strategy} we
introduce the model and explain how we analyze it. To validate our
results, we compare different versions of a deterministic continuum
model to stochastic computer simulations of filamentous
networks. Subsequently, the impact of different piecewise straight
(\textit{linear}) obstacle shapes on the resulting network orientation
patterns is analyzed in \sec{linear_obstacle_shape}, while specific
examples for curved (\textit{nonlinear}) obstacle shapes are the
subject of \sec{nonlinear_obstacle_shape}. Our main result is that the
competition between two fundamentally different network architectures
persists for finite-sized, piecewise linear and curved obstacles.

\section{Model definition}
\label{modeling_strategy}

\begin{figure}[htp]
\centering
\includegraphics[width=0.9\textwidth]{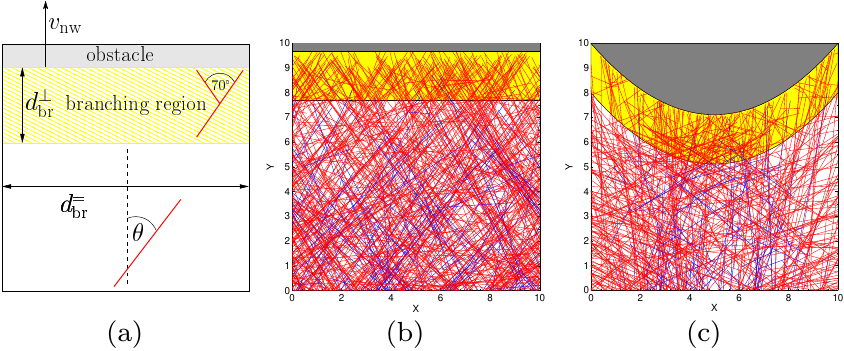}
\caption{Model of a growing actin network behind a rigid obstacle. (a)
  Sketch of the setup. Branching can occur only in the yellow
  region. The main quantity of interest is the filament orientation angle
  $\theta$ relative to the surface normal. (b) Snapshot of a
  stochastic simulation based on the reactions of individual filaments
  for straight (\textit{linear}) obstacle shape. (c) Snapshot of
  network growth behind a curved (\textit{nonlinear}) obstacle
  shape. The network is growing in two dimensions. Red filaments are
  actively growing at their barbed end in direction $\theta$. Blue filaments have been capped
  and will eventually be outgrown by the network and leave the box at
  the bottom. Growth of the fastest filaments is stalled by the obstacle.}
\label{figure1}
\end{figure}

Motivated by established biological observations, in our model we assume Arp2/3 to nucleate daughter filaments
from preexisting mother filaments at a characteristic relative branching angle
around $70^\circ$ \cite{mullins1998,svitkina1999,ydenberg2011}.
The exact value varies
with biological system and analysis methods (for example, a recent
value from electron tomography data for the lamellipodium of
fibroblasts is $73\pm8^\circ$ \cite{vinzenz2012}), but is not essential
for our theory. Although the exact mechanism for the activation of
Arp2/3-mediated branching is not yet well established and subject of
current research, it is generally accepted that Arp2/3 is active only
near the obstacle, where it is activated by nucleation
promoting factors (NPFs) such as the Wiskott–Aldrich syndrome protein
(WASP) or the bacterial actin assembly-inducing protein (ActA)
\cite{beltzner2008,ti2011,xu2012}. As we are interested in actin
network architecture in close proximity to the surface of a propelled
obstacle, we will focus on actin dynamics within the first few ten
nanometers from the obstacle surface in which filament branching,
capping and barbed end polymerization are expected to dominate actin
dynamics, while filament depolymerization and decapping can be
neglected \cite{schafer1996}. \fig{figure1}(a) sketches the geometrical arrangement
studied here. The yellow {\it branching region} within a vertical
distance $d_{\rm br}^\perp$ from the obstacle surface indicates the
domain in which filament bound Arp2/3 is able to interact with NPFs, which themselves are active only close to the surface. Thus filament branching occurs in this branching region. Once
filament barbed ends have left this domain due to their retrograde flow they are not able to
nucleate new daughter filaments anymore and will eventually be
outgrown by the bulk network. As those filaments subsequently do not impact the orientation distribution at the leading edge, we do not explicitly account for their further fate anymore. If the lateral extension $d_{\rm br}^=$
of the obstacle is large compared to the width $d_{\rm br}^\perp$ of
the branching region, it is possible to neglect the process of
filament barbed ends growing out of the branching region
horizontally. However, for relatively small obstacles, for instance
viral pathogens, the lateral dimension of the propelled particle
$d_{\rm br}^=$ can become comparable to its vertical extension $d_{\rm
  br}^\perp$ and the finite lateral size of the branching region has
to be taken into account. Our main quantity of interest is the
filament orientation angle $\theta$ relative to the surface normal.

In the following, we will validate our theoretical approach by
comparing results from two complementary implementations, namely
stochastic computer simulations and a rate equation approach, which
have been introduced before for flat obstacles of large lateral size
\cite{weichsel2010_2}. Motivated by the flat nature of the
lamellipodium and also for computational simplicity, our modeling is
restricted to two dimensions. Representative snapshots of such
simulations for flat and curved obstacle geometries are given in
\fig{figure1}(b) and (c), respectively. Briefly, we simulate a network of infinitely stiff rods. This simplification is justified as the reaction kinetics that eventually determine the stable orientation distribution in steady state are active within a narrow branching region along the obstacle only. Because filament segments spanning over this region are very short compared to their persistence length (nanometers versus micrometers), local bending undulations are of minor importance here. The remainder of the filaments is considered to be entangled in the bulk actin network, which effectively acts as a base for the protruding filaments. However, filaments embedded in semidilute actin solutions have been found to exhibit substantial bending fluctuations in the past \cite{kaes1994,dichtl1999} and therefore we effectively account for a finite uncertainty in filament orientation by incorporating a distribution of relative branching angles between mother and daughter filaments in the model as is discussed below. Each uncapped filament is growing
deterministically at its barbed end with a fixed velocity $v_{\rm
  fil}$. Thus we implicitly assume a constant density of actin
monomers at the leading edge. Polymerization is quantized such that
filaments extent by one building block of length $\delta_{\rm fil}$
per unit time. Apart from branching, individual filaments do not
interact and hence the local filament density does not alter polymerization. 
Filament barbed ends within the branching region close to
the leading edge are possible candidates for stochastic branching and
capping events.  While capping is assumed to be a first order reaction
in the number of actin barbed ends in the reaction zone, for branching
we assume zeroth order. This is motivated by the expectation that the
supply of activated Arp2/3 is strongly limited by the
availability of NPFs at the leading edge and thus an effective zeroth
order branching rate emerges for sufficiently high filament density or
low capping rate \cite{pollard2000,weichsel2012_2}. In the opposite limit of low filament density, the branching reaction is expected to become first order in the number of filaments; this effectively yields an autocatalytic description \cite{carlsson2003}, that is stationary only at one unique network growth velocity. Therefore in this limit, transitions in the filament orientation distribution are not accessible. However, it has been shown that both, first and zeroth order branching models, can be incorporated within a unified theoretical framework and that the crossover from one regime to the other does not have a direct impact on filament orientation in steady state \cite{weichsel2012_2}. As actin networks in most experimental setups are growing against a finite load of the obstacle and non-constant force-velocity curves are observed, the zeroth order branching regime is expected to dominate the branching kinetics eventually.  

The orientation of new filament branches is chosen from a
normalized linear combination of two Gaussian probability
distributions for the angle between mother and daughter filament with
means at $\pm70^\circ$ and standard deviation $5^\circ$. Red filaments
in the illustration are actively growing, while blue filaments
indicate capped barbed ends that neither branch nor polymerize anymore
and will eventually be outgrown by the bulk network and leave the
simulation box at the bottom. In \fig{figure1}(b) the top boundary of
the yellow branching region defines a rigid obstacle, which excludes
volume and thus prevents polymerization above its boundary. As a
consequence the fastest filaments which are growing in close to
vertical direction are stalled by the obstacle. The vertical velocity
of the obstacle is a parameter of the model and set to a constant
value $v_{\rm nw}$ directly. This growth velocity should be regarded as the effective outcome in steady state, determined by the details of filament-obstacle interactions (including regulation by biochemical factors). Steady state growth is possible within the velocity range, $0 < v_{\rm nw} < v_{\rm fil}$. Under these conditions a well defined filament number and orientation distribution evolves.

As a powerful analytical alternative to the stochastic framework
introduced above, we also develop a deterministic rate equation for
dendritic network growth based on earlier approaches of this kind
\cite{maly2001,carlsson2003,weichsel2010_2}. The evolution of the
orientation distribution $N(\theta,t)$ of uncapped filament barbed
ends integrated over the whole branching region evolves in time by
branching and capping events and by filaments growing out of the
branching region. This translates into the following ordinary
differential equation:
\begin{equation}
\frac{\partial N(\theta,t)}{\partial t} = \underbrace{\hat{k}_{\rm b}\int \mathcal{W}(\theta,\theta') N(\theta',t) \,\mbox{d}\theta'}_{\text{branching}} 
- \underbrace{k_{\rm c} N(\theta,t)}_{\text{capping}} - \underbrace{k_{\rm gr}(\theta, v_{\rm nw}) N(\theta,t)}_{\text{outgrowth}}\,.
\label{system_dgl_fiber_number} 
\end{equation}
Here again, capping is assumed to be a first order reaction
proportional to the number of existing filaments with the
proportionality constant, a reaction rate $k_{\rm c}$. The probability
of branching at a given angle is determined by the weighting factor
distribution, $\mathcal{W}(\theta,\theta')$, which is modeled as a
linear combination of Gaussians with maxima at branching angles
$\pm70^\circ$ between filaments and standard deviation
$\sigma=5^\circ$. $\theta$ is the filament orientation angle measured
relative to the vertical direction. The branching reaction is
independent of the absolute number of existing filament barbed ends in
this model (i.e. a zeroth order reaction). Hence, the branching rate
$\hat{k}_{\rm b}$ is normalized by the total number of new
filament ends,
 \begin{equation}
\hat{k}_{\rm b}=\frac{k_{\rm b}}{\mathcal{W}_{\rm tot}}\ ,\ \mathcal{W}_{\rm tot}=
\int\int \mathcal{W}(\theta,\theta') N(\theta',t) \,\mbox{d}\theta'\,\mbox{d}\theta\,.
\label{lamellipodium_normalization}
\end{equation}
The reaction rate $k_{\rm b}$ indicates the number of new branches per
unit time.  Most importantly for the specific scope of this work,
obstacle velocity and shape enter \eq{system_dgl_fiber_number} via the
outgrowth rate $k_{\rm gr}$. As filaments are polymerizing with a
constant velocity $v_{\rm fil}$ in their individual direction, some
filaments will not be able to keep up with the moving obstacle and thus
leave the branching region. The precise expression for the rate of
outgrowth $k_{\rm gr}$ strongly depends on obstacle geometry and
alters the resulting steady state orientation distributions as will be
analyzed in detail in the following.

\section{Linear obstacle shape}
\label{linear_obstacle_shape}

We begin our analysis with linear obstacle shapes as illustrated in
\fig{figure1}(a) and (b). The obstacle (and therefore the leading edge
of the network) moves with a constant velocity $v_{\rm nw}$ towards
the top. Our region of interest, the branching region, extends
vertically to a distance $d_{\rm br}^\perp$ from the leading edge of
the network. The lateral obstacle width is indicated by $d_{\rm
  br}^=$.

\subsection{Flat large obstacle}
\label{flat_obstacle_shape}

For obstacles that are laterally widely extended, for instance the
leading edge at the front of the lamellipodium in a migrating cell, we
have $d_{\rm br}^= \gg d_{\rm br}^\perp$ and thus horizontal filament
outgrowth can be neglected locally. Mathematically, this means that we
can use periodic boundary conditions in the horizontal direction. This
case has been analyzed before in a similar manner as done below for
finite-sized and curved obstacles \cite{weichsel2010_2}, and therefore
we recapitulate the most important results as a reference case.  In
this simple case, outgrowth of filaments from the branching region can
only occur in negative normal direction. While the network grows with
velocity $v_{\rm nw}$ in positive normal direction, single filaments
grow with velocity $v_{\rm fil}$ in their individual direction. The projected polymerization
velocity depends on the filament orientation, as
\begin{equation}
 v_{\rm fil}^\perp (\theta)= v_{\rm fil} \cos \theta \,.
\label{lamellipodium_perp_fil_vel}
\end{equation}
Filaments having a larger absolute orientation than the critical angle
\begin{equation}
\theta_{\rm c}=\arccos \left( \frac{v_{\rm nw}}{v_{\rm fil}} \right) \,,
\label{lamellipodium_crit_angle}
\end{equation}
are thus not able to keep up with the speed of the obstacle and are subject to outgrowth with a rate
\begin{equation}
{k_{\rm gr}^{\perp}}(\theta, v_{\rm nw})=
\begin{cases} 
0 & \text{if $\vert\theta\vert \leq \theta_{\rm c}$}\\
\frac{v_{\rm nw}-v_{\rm fil} \cos{\theta}}{(d_{\rm br}^\perp/2)} &\text{if $\vert\theta\vert > \theta_{\rm c}$}
\end{cases}\ .
\label{lamellipodium_fil_lifetime}
\end{equation}
The factor $2$ results from the assumption that new filaments
branch-off from existing filaments on average in the center of the
branching region of vertical width $d_{\rm br}^\perp$. Outgrowth
vanishes for filaments with an orientation smaller than a critical
angle $\theta_{\rm c}$ and from there increases to its maximum at
$\vert\theta\vert = 180^{\circ}$.

We solve \eq{system_dgl_fiber_number} numerically by introducing 360
angle bins and iterating the equations until a steady state is
achieved, which then can be compared to the results of the
stochastic computer simulations. In order to achieve a deeper
understanding, we also introduced a coarse-grained
version of \eq{system_dgl_fiber_number} that can be treated
analytically. In \eq{system_dgl_fiber_number}, the number of filament
ends in the branching region with angles between $\theta$ and
$\theta+d\theta$ is given by $N(\theta,t)d\theta$. By extending this
integration to sufficiently large ($\Delta\theta=35^\circ$) angle
bins
\begin{equation}
N_{\bar{\theta}}=\int\limits_{\bar{\theta}-\Delta\theta/2}^{\bar{\theta}+\Delta\theta/2} N(\theta',t) \,\mbox{d}\theta'\,,
\end{equation}
and further assuming that branching is restricted to pairs of angle
bins with a relative angle difference of $70^\circ$, a system of five coupled ordinary
differential equations results:
\begin{eqnarray}
\frac{\partial N_{-70^\circ}}{\partial t} & = & \frac{1}{2} \hat{k}_{\rm b} N_{0^\circ} - \left(k_{\rm c} +{k_{\rm gr}^{\perp}}(70^\circ)\right)N_{-70^\circ}
\label{dgl_N-70}\\ 
\frac{\partial N_{-35^\circ}}{\partial t} & = & \frac{1}{2} \hat{k}_{\rm b} N_{+35^\circ} - \left(k_{\rm c} +{k_{\rm gr}^{\perp}}(35^\circ)\right) N_{-35^\circ}
\label{dgl_N-35}\\ 
\frac{\partial N_{0^\circ}}{\partial t} & = & \frac{1}{2} \hat{k}_{\rm b} \left(N_{-70^\circ}+N_{+70^\circ} \right) - k_{\rm c} N_{0^\circ} \label{dgl_N0}\\ 
\frac{\partial N_{+35^\circ}}{\partial t} & = & \frac{1}{2} \hat{k}_{\rm b} N_{-35^\circ} - \left(k_{\rm c} +{k_{\rm gr}^{\perp}}(35^\circ)\right) 
N_{+35^\circ} 
\label{dgl_N+35}\\ 
\frac{\partial N_{+70^\circ}}{\partial t} & = & \frac{1}{2} \hat{k}_{\rm b} N_{0^\circ} - \left(k_{\rm c} +{k_{\rm gr}^{\perp}}(70^\circ)\right)N_{+70^\circ}\,, 
\label{dgl_N+70}
\end{eqnarray}
with 
\begin{equation*}
\hat{k}_{\rm b}=\frac{k_{\rm b}}{\mathcal{W}_{\rm tot}}=\frac{k_{\rm b}}{N_{-70^\circ}+N_{-35^\circ}+N_{0^\circ}+N_{+35^\circ}+N_{+70^\circ}}\,.
\end{equation*}
Here we have also assumed that branching of filaments with
orientations $\vert \theta \vert > 87.5^\circ$ can be neglected. The
five equations are symmetric around $0^\circ$, and thus only three of
them are independent. Nonlinearity and the coupling of all five
equations is introduced by the branching term due to the zeroth order
branching reaction.

By algebraically solving \eq{dgl_N-70})--(\eq{dgl_N+70} for the
stationary state, we obtain two physically meaningful solutions. The
first solution,
\begin{eqnarray}
\begin{array}{rcl}
N_{-70^\circ}^{\rm ss35} & = & 0 \\
N_{-35^\circ}^{\rm ss35} & = & k_{\rm b} \frac{1}{4 (k_{\rm c} + {k_{\rm gr}^{\perp}}(35^\circ))}\\
N_{0^\circ}^{\rm ss35} & = &  0 \\
N_{+35^\circ}^{\rm ss35} & = & k_{\rm b} \frac{1}{4 (k_{\rm c} + {k_{\rm gr}^{\perp}}(35^\circ))}\\
N_{+70^\circ}^{\rm ss35} & = & 0\,,
\end{array}
\label{steadystate_sol_35}
\end{eqnarray}
represents a dominant $\pm 35$ degrees orientation distribution in the steady state while the second solution,
\begin{eqnarray}
\begin{array}{rcl}
N_{-70^\circ}^{\rm ss70} & = & k_{\rm b} \frac{k_{\rm c} + {k_{\rm gr}^{\perp}}(70^\circ) - \sqrt{2 k_{\rm c} \left( k_{\rm c} + {k_{\rm gr}^{\perp}}(70^\circ) \right)}}
{2 \left({k_{\rm gr}^{\perp}}^2(70^\circ) - k_{\rm c}^2 \right)}\\
N_{-35^\circ}^{\rm ss70} & = & 0\\
N_{0^\circ}^{\rm ss70} & = &  k_{\rm b} \frac{1 - \sqrt{\frac{k_{\rm c} + {k_{\rm gr}^{\perp}}(70^\circ)}{2 k_{\rm c}}}}
{k_{\rm c} - {k_{\rm gr}^{\perp}}(70^\circ)}\\
N_{+35^\circ}^{\rm ss70} & = & 0\\
N_{+70^\circ}^{\rm ss70} & = & k_{\rm b} \frac{k_{\rm c} + {k_{\rm gr}^{\perp}}(70^\circ) - \sqrt{2 k_{\rm c} \left( k_{\rm c} + {k_{\rm gr}^{\perp}}(70^\circ) \right)}}
{2 \left({k_{\rm gr}^{\perp}}^2(70^\circ) - k_{\rm c}^2 \right)} \,,
\end{array}
\label{steadystate_sol_70}
\end{eqnarray}
corresponds to a competing $+70/0/\!\!-\!\!70$ pattern.

The stability of these two fixed points can be analyzed with respect
to the parameters $k_{\rm b}$, $k_{\rm c}$, $k_{\rm gr}^{\perp}(35^\circ)$, ${k_{\rm gr}^{\perp}}(70^\circ)$ and $d_{\rm br}^\perp$ using linear stability analysis. The
result of this analysis is independent of the branching rate $k_{\rm
  b}$ and therefore this parameter has no influence on the stability
of the system. According to \eq{steadystate_sol_35} and \eq{steadystate_sol_70}, however, the total number of filaments in steady state is proportional to $k_{\rm b}$ and therefore this parameter can be used to adjust the model to experimentally measured densities. The two parameters ${k_{\rm gr}^{\perp}}(35^\circ)$ and
$k_{\rm gr}^{\perp}(70^\circ)$ are not independent, but rather both of
them are determined by the obstacle velocity $v_{\rm nw}$ as given in
\eq{lamellipodium_fil_lifetime}. In the following, we will omit the
ill-defined cases $k_{\rm c}={k_{\rm gr}^{\perp}}(35^\circ)=0$ and
$k_{\rm c}={k_{\rm gr}^{\perp}}(70^\circ)=0$, for which the filament
number diverges. We find that the stability of both fixed
points changes when
\begin{equation}
k_{\rm c}{k_{\rm gr}^{\perp}}(70^\circ)=k_{\rm c}^2+4 k_{\rm c} {k_{\rm gr}^{\perp}}(35^\circ) + 2 {k_{\rm gr}^{\perp}}^2(35^\circ)
\label{eigenvalue35_zero}
\end{equation}
and that either one is asymptotically stable, while the other is a
saddle. Thus the simple model suggests that exactly two possible
network architectures exist with mutually exclusive stability.

\eq{eigenvalue35_zero} can now be used to obtain the respective
network velocity $v_{\rm nw}$ of the transition. For a critical angle $\theta_{\rm c}
\geq 70^\circ$ (i.e. $k_{\rm gr}^{\perp}(70^\circ)={k_{\rm
    gr}^{\perp}}(35^\circ)=0$), \eq{eigenvalue35_zero} is never
satisfied (note that $k_{\rm c} > 0$). For $35^\circ \leq \theta_{\rm c} <
70^\circ$ (i.e. ${k_{\rm gr}^{\perp}}(70^\circ)>0 \wedge{k_{\rm
    gr}^{\perp}}(35^\circ)=0$),
\begin{equation}
v_{\rm nw} = \frac{1}{2} k_{\rm c} d_{\rm br}^\perp + v_{\rm fil} \cos(70^\circ) \,
\textnormal{, for } 35^\circ \leq \theta_{\rm c} < 70^\circ\ .
\label{vgr_sol1}
\end{equation}
satisfies \eq{eigenvalue35_zero}.  For network velocities with a
critical angle $\theta_{\rm c}<35^\circ$ (i.e. ${k_{\rm
    gr}^{\perp}}(70^\circ)>{k_{\rm gr}^{\perp}}(35^\circ)>0$), two
solutions emerge,
\begin{eqnarray}
\begin{array}{rcl}
v_{\rm nw\,1,2} & = & \frac{1}{8} \left( - 3 k_{\rm c} d_{\rm br}^\perp + 8 v_{\rm fil} \cos(35^\circ) \right) \\
       &   & \pm \frac{1}{8} \sqrt{k_{\rm c} d_{\rm br}^\perp \left( k_{\rm c} d_{\rm br}^\perp + 16 v_{\rm fil} \cos(35^\circ) - 16 v_{\rm fil} \cos(70^\circ) \right) }\,,\\
       &   & \textnormal{for } \theta_{\rm c}<35^\circ\,. \\
\end{array}
\label{vgr_sol2}
\end{eqnarray}
It turns out that solution \eq{vgr_sol1} is valid for network bulk
velocities $v_{\rm fil} \cos(35^\circ) \geq v_{\rm nw} > v_{\rm fil}
\cos(70^\circ)$, while for solution \eq{vgr_sol2} to be valid $v_{\rm
  nw} > v_{\rm fil} \cos(35^\circ)$ has to be satisfied. This is never
the case for the negative square-root in \eq{vgr_sol2} and so we can
neglect this solution in the following. In order to further simplify
our equations, we define the reference velocity $u_{\rm c} \equiv
k_{\rm c} d_{\rm br}^\perp$ resulting from the capping rate.
\eq{vgr_sol1} and \eq{vgr_sol2} then become
\begin{equation}
\frac{v_{\rm nw}}{v_{\rm fil} } = \frac{1}{2} \frac{u_{\rm c}}{v_{\rm fil} } + \cos(70^\circ) \,,
\label{vgr_sol1_eff_par}
\end{equation}
and,
\begin{eqnarray}
\begin{array}{rcl}
\frac{v_{\rm nw}}{v_{\rm fil}} & = & \frac{1}{8} \left( - 3 \frac{u_{\rm c}}{v_{\rm fil}} + 8 \cos(35^\circ) \right) \\
    &   & + \frac{1}{8} \sqrt{\frac{u_{\rm c}}{v_{\rm fil}} \left( \frac{u_{\rm c}}{v_{\rm fil}} + 16 \cos(35^\circ) - 16 \cos(70^\circ) \right) }\,,
\end{array}
\label{vgr_sol2_eff_par}
\end{eqnarray}
respectively. Thus the capping rate emerges as an essential
parameter through the effective velocity $u_{\rm c}$.

In \fig{figure2}(a) the stability diagram from the analytical
analysis is illustrated, showing the regions in which either the $\pm35$ degree
distribution is asymptotically stable and the $+70/0/\!\!-\!\!70$ pattern
is a saddle or vice versa. In order to compare these results to the
full numerical solution of \eq{system_dgl_fiber_number}, we define the
relative difference of filaments in the angle bin around $0^\circ$ to
the one around $35^\circ$ as an appropriate order parameter in steady
state,
\begin{equation}
\mathcal{O}=\frac{N_{0^\circ}-N_{35^\circ}}{N_{0^\circ}+N_{35^\circ}}=\left[-1,+1\right]\,.
\label{lamellipodium_order_paramer}
\end{equation}
For a perfect $+70/0/\!\!-\!\!70$ distribution this parameter will
approach $+1$, while for the competing $\pm 35$ pattern it will
approach $-1$. The transition between the two patterns is defined when
the order parameter changes its sign. According to this definition of
the transition point, it is now possible to numerically solve the
continuum model \eq{system_dgl_fiber_number} and to classify each
observed stationary state as one of the two phases. These results are
presented as a phase diagram in \fig{figure2}(d) and were obtained for
different branching rates $k_{\rm b}$, which indeed has no significant
influence as expected from the analytical considerations. In the
solution of the full continuum model the capping rate only has a very
limited influence on the stripe-like pattern of the phase
diagram. Contrarily, in the phase diagram resulting from linear
stability analysis (\fig{figure2}(a)), the $\pm35$ pattern vanishes at
large capping rate. This artifact of the reduced rate equation model
is introduced by the assumption that filaments with an orientation
larger than $87.5^\circ$ do not branch.  As the full rate equation
does not share this assumption, it does not predict the elimination of
the $\pm35$ pattern for large capping rate.

\subsection{Flat finite-sized obstacle}
\label{nonperiodic_horizontal_obstacle_boundaries}

For flat, but finite-sized obstacle shape, filaments might also leave
the branching region horizontally. Therefore a second outgrowth rate
needs to be incorporated into the rate equation
(\eq{system_dgl_fiber_number}):
\begin{equation}
k_{\rm gr}^=(\theta, v_{\rm nw})=
\begin{cases} 
\frac{v_{\rm nw} \tan{\theta}}{(d_{\rm br}^=/2)} & \text{if $\vert\theta\vert \leq \theta_{\rm c}$}\\
\frac{v_{\rm fil} \sin{\theta}}{(d_{\rm br}^=/2)} &\text{if $\vert\theta\vert > \theta_{\rm c}$}
\end{cases}
\, .
\label{particle_lat_outgrowth_rate}
\end{equation}
We find that for this scenario, it is still possible to analyze the stability of
the fixed points of \eq{dgl_N-70}--\eq{dgl_N+70}. The only difference
that accounts for the change in obstacle geometry at this point is
that a combination of outgrowth rates $k_{\rm gr}(\theta) = k_{\rm
  gr}^\perp(\theta) + \vert k_{\rm gr}^=(\theta) \vert$ replaces the
term for exclusively orthogonal outgrowth $k_{\rm gr}^\perp(\theta)$
considered in the previous section. For filaments growing in the
$N_{0^\circ}$ angle bin, both outgrowth rates vanish, because these
filaments are growing at a high enough velocity parallel to the lateral
boundaries of the obstacle.

The same arguments as used before also apply for the stability
analysis of the steady states here. Hence, we can begin with
\eq{eigenvalue35_zero} and evaluate the stability of the two steady
states according to the adjusted outgrowth rate. To determine the
transitions, we will follow a similar strategy as before. Starting
from small network velocities $v_{\rm nw}$, we will treat the
different possibilities for the outgrowth rates given in
\eq{lamellipodium_fil_lifetime} and \eq{particle_lat_outgrowth_rate}
in a case-by-case analysis.

For $\theta_{\rm c}\geq70^\circ$, the orthogonal outgrowth rates
vanish, as all relevant filament orientations grow faster towards the
top than the obstacle and are slowed down to $v_{\rm nw}$ in this
direction. The outgrowth rates for the different orientation bins
therefore read,
\begin{equation}
k_{\rm gr}^{\perp}(\theta)=0 \text{ and } k_{\rm gr}^{=}(\theta)=\frac{v_{\rm nw} \tan{\theta}}{(d_{\rm br}^=/2)} \text{ for } \theta=[35^\circ,70^\circ]\,.
\label{particle_outgrowth_slow}
\end{equation}
Inserting $k_{\rm gr} = k_{\rm gr}^\perp + k_{\rm gr}^= $ in
\eq{eigenvalue35_zero} and omitting again the ill-defined cases of
$k_{\rm b}=0$, $k_{\rm c}=0$ and $d_{\rm br}^= = 0$ here (and
additionally $d_{\rm br}^\perp = 0$ in the following), it can be shown
that \eq{eigenvalue35_zero} is never satisfied.

Increasing the network velocity to the point where $35^\circ \leq \theta_{\rm c} < 70^\circ$ leads to outgrowth rates,
\begin{equation}
k_{\rm gr}^{\perp}(35^\circ)=0 \text{ and } k_{\rm gr}^{\perp}(70^\circ)=\frac{v_{\rm nw} - v_{\rm fil} \cos(70^\circ)}{(d_{\rm br}^\perp/2)} \,,
\label{particle_outgrowth_orth_middle}
\end{equation}
in orthogonal direction according to \eq{lamellipodium_fil_lifetime} and,
\begin{equation}
k_{\rm gr}^{=}(35^\circ)=\frac{v_{\rm nw} \tan(35^\circ)}{(d_{\rm br}^=/2)} \text{ and } k_{\rm gr}^{=}(70^\circ)=\frac{v_{\rm fil} \sin(70^\circ)}{(d_{\rm br}^=/2)} \,,
\label{particle_outgrowth_lat_middle}
\end{equation}
in the lateral direction according to \eq{particle_lat_outgrowth_rate}. These rates yield a quadratic equation for the network velocities $v_{\rm nw}$ that satisfy \eq{eigenvalue35_zero},
\begin{equation}
\begin{array}{rcl}
0 & = & v_{\rm nw}^2 \left[ 8 \tan^2(35^\circ) \frac{1}{{u_{\rm c}}^2 {r_{\rm br}}^2} \right] + v_{\rm nw} \left[ 8 \tan(35^\circ) \frac{1}{u_{\rm c} r_{\rm br}} - 2 \frac{1}{u_{\rm c}} \right]\\
  &   & + \left[ 1  - 2 v_{\rm fil} \sin(70^\circ) \frac{1}{u_{\rm c} r_{\rm br}} + 2 v_{\rm fil} \cos(70^\circ) \frac{1}{u_{\rm c}} \right] \,, \\
  &   & \textnormal{for } 35^\circ \leq \theta_{\rm c} < 70^\circ \,.
\end{array}
\label{particle_vnw_sol_middle_eff_par}
\end{equation}
Here, the reference velocity, $u_{\rm c}=k_{\rm c} d_{\rm br}^\perp$,
together with a new parameter for the length scale ratio in lateral
and orthogonal direction, $r_{\rm br} \equiv d_{\rm br}^= / d_{\rm
  br}^\perp$, are identified to determine stability.

For a critical angle $\theta_{\rm c} < 35^\circ$, the adjusted outgrowth rates read,
\begin{equation}
k_{\rm gr}^{\perp}(\theta) = \frac{v_{\rm nw}-v_{\rm fil} \cos{\theta}}{(d_{\rm br}^\perp/2)} \text{ and } k_{\rm gr}^{=}(\theta)=\frac{v_{\rm fil} \sin{\theta}}{(d_{\rm br}^=/2)} \text{ for } \theta=[35^\circ,70^\circ]\,.
\label{particle_outgrowth_fast}
\end{equation}
These rates together with \eq{eigenvalue35_zero} and the effective
parameters $u_{\rm c}$ and $r_{\rm br}$ can be simplified to a second
quadratic equation defining the network velocities at the transitions,
\begin{equation}
\begin{array}{rcl}
0 & = &  v_{\rm nw}^2 \left[8 \frac{1}{u_{\rm c}^2} \right] +  v_{\rm nw} \left[ 6 \frac{1}{u_{\rm c}} - 16 v_{\rm fil} \cos(35^\circ) \frac{1}{u_{\rm c}^2} + 16  v_{\rm fil} \sin(35^\circ) \frac{1}{u_{\rm c}^2 r_{\rm br}} \right] \\
  &   & + \left[ 1 + 2 v_{\rm fil} \cos(70^\circ) \frac{1}{u_{\rm c}} - 2 v_{\rm fil} \sin(70^\circ) \frac{1}{u_{\rm c} r_{\rm br}}- 8 v_{\rm fil} \cos(35^\circ) \frac{1}{u_{\rm c}} + 8 v_{\rm fil} \sin(35^\circ) \frac{1}{u_{\rm c} r_{\rm br}} \right. \\
  &   & \hspace{0.5cm} + \left. 8 v_{\rm fil}^2 \cos^2(35^\circ) \frac{1}{u_{\rm c}^2} - 16 v_{\rm fil}^2 \cos(35^\circ) \sin(35^\circ) \frac{1}{u_{\rm c}^2 r_{\rm br}} + 8 v_{\rm fil}^2 \sin^2(35^\circ) \frac{1}{u_{\rm c}^2 r_{\rm br}^2} \right] \,, \\
  &   & \textnormal{for } \theta_{\rm c} < 35^\circ \,.
\end{array}
\label{particle_vnw_sol_fast_eff_par}
\end{equation}

\begin{figure}[htp]
\centering
\includegraphics[width=0.9\textwidth]{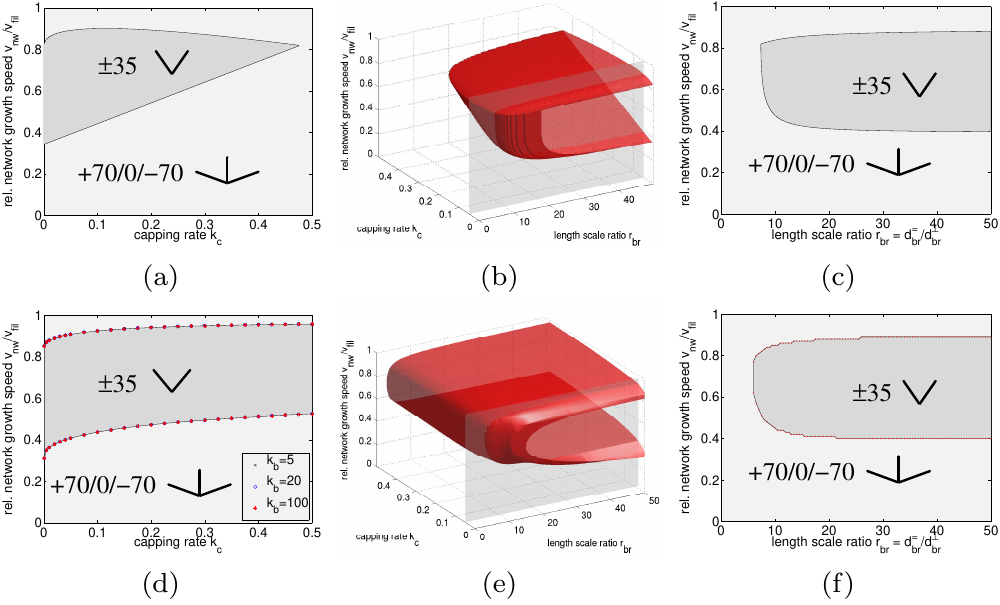}
\caption{Phase diagram for flat finite-sized obstacles. (a-c)
  Analytical results from linear stability analysis of the simplified
  continuum model with 5 angle bins. (d-f) Full numerical solution of
  the rate equation model with $360$ angle bins. In (a) and (d), 2D
  slices through the 3D hypersurface are plotted within the $k_{\rm
    c}$-$v_{\rm nw}$ plane in the limit of $r_{\rm br}\rightarrow
  \infty$, which corresponds to periodic boundary conditions or very
  large objects. (b) and (e) show the hyperplane of transition points between the two filament patterns in the full 3D parameter space. In (c) and (f), slice plots in the $r_{\rm
    br}$-$v_{\rm nw}$ plane are shown along the transparent gray plane
  in (b) and (e) at $k_{\rm c}=0.05$.}
\label{figure2}
\end{figure}

According to \eq{particle_vnw_sol_middle_eff_par} and
\eq{particle_vnw_sol_fast_eff_par}, the regions of stability for the
two different stationary orientation patterns in parameter space are
illustrated in \fig{figure2}(b). As the finite obstacle width
introduces an additional independent parameter $r_{\rm br}$, the full
parameter space is now three dimensional. Network growth velocities
$v_{\rm nw}$ that fulfill \eq{particle_vnw_sol_middle_eff_par}
together with the condition $v_{\rm fil} \cos(35^\circ) \geq v_{\rm
  nw} > v_{\rm fil} \cos(70^\circ)$ or
\eq{particle_vnw_sol_fast_eff_par} at $v_{\rm nw} > v_{\rm fil}
\cos(35^\circ)$ are identified as transition points between the
different orientation distributions in the parameter space spanned by
$k_{\rm c}$, $r_{\rm br}$ and $v_{\rm nw}$. In the limit of large
length scale ratio, $r_{\rm br} \rightarrow \infty$ (i.e. $d_{\rm
  br}^= \gg d_{\rm br}^\perp$), lateral outgrowth can be neglected and
the phase diagram approaches the results for periodic boundary
conditions as given in \fig{figure2}(a). For relatively small $r_{\rm
  br}\lesssim7$, the lateral outgrowth of intermediate filament
orientations at around $\pm35^\circ$ is sufficiently large to prevent
this orientation pattern from being stable in this model, independent
of the growth velocity (cf. \fig{figure2}(c)).

Using the order parameter \eq{lamellipodium_order_paramer}, we
numerically sampled the parameter space according to
\eq{system_dgl_fiber_number} with adjusted outgrowth. The isosurface
$\mathcal{O}(v_{\rm nw}, k_{\rm c}, r_{\rm br})=0$ is extracted from
the three dimensional data and shown in \fig{figure2}(e). In the limit
of large length scale ratio, the results coincide well with the case of
periodic boundary conditions (cf. \fig{figure2}(d)). It
is also confirmed that for small values of $r_{\rm br}$ the $\pm35$
pattern is not stable anymore at intermediate velocities (cf. \fig{figure2}(f)).
Thus the overall agreement between the simple analytical model
and the full numerical solution of the rate equation approach
is surprisingly good.

\begin{figure}[htp]
\centering
\includegraphics[width=0.9\textwidth]{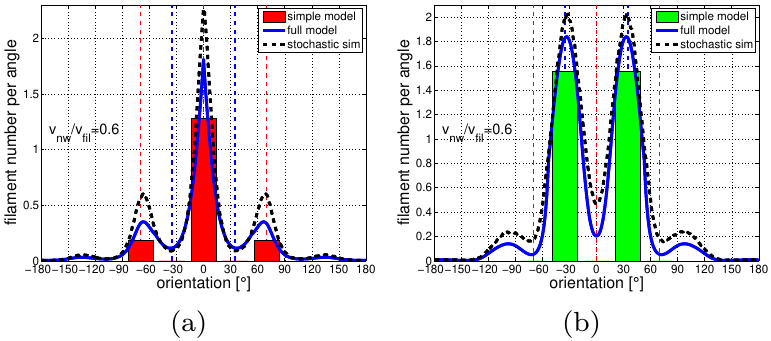}
\caption{Comparison of the steady state orientation distributions for actin
  networks growing in the tail of a flat 
  obstacle of finite size at $v_{\rm nw}/v_{\rm fil}=0.6$, $k_{\rm b}=20$, $k_{\rm
    c}=0.05$, $d_{\rm br}^\perp=2\delta_{\rm
    fil}$ and (a) $r_{\rm br}=3$, (b) $r_{\rm br}=20$. The results
  were obtained according to three different methods, the simplified
  continuum model (bars), the full continuum model (blue solid line)
  and stochastic simulations (black dashed line).}
\label{figure3}
\end{figure}

\fig{figure3} compares filament orientation distributions obtained in
steady state from stochastic network simulations, the full continuum
model and the simplified continuum model. The parameters of the two
cases shown in (a) and (b) do not differ in their network velocity
$v_{\rm nw}$, but rather only in the obstacle geometry, in this case
given by length scale ratio $r_{\rm br}=d_{\rm br}^=/d_{\rm
  br}^\perp$. For small $r_{\rm br}=3$ at intermediate velocities, the
$\pm35$ pattern of the network is not stable anymore and rather the
network organizes in the alternative $+70/0/\!\!-\!\!70$ distribution
(cf. \fig{figure3}(a)). For larger ratio $r_{\rm br}=20$, the network
organizes in the $\pm35$ degree distribution
(cf. \fig{figure3}(b)). In the stochastic simulations, outgrowth
rates are not explicitly incorporated, but rather emerge as a direct
consequence of the obstacle geometry. Thus the computer simulations
nicely validate both, the full and reduced continuum approaches,
and all three seem to capture the essential physical mechanisms
determining network structure.

\subsection{Obstacle with tilted straight contour}

\begin{figure}[htp]
\centering
\includegraphics[width=0.9\textwidth]{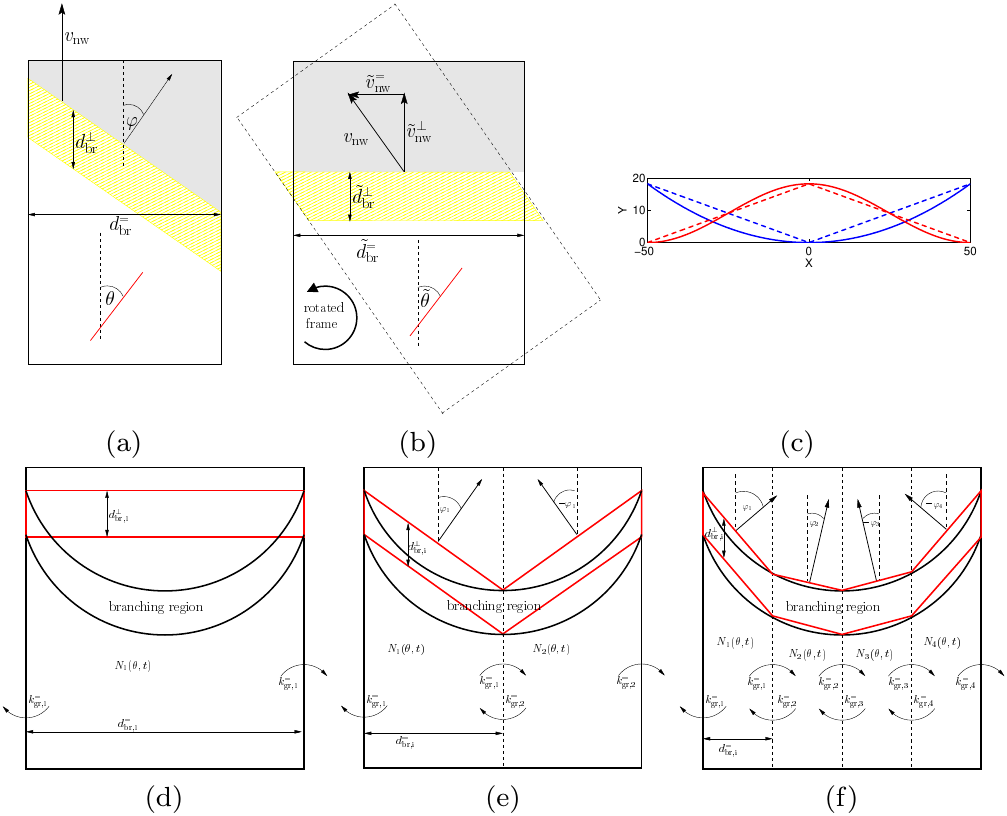}
\caption{(a) and (b) Within a rotated coordinate frame, the continuum
  model (\eq{system_dgl_fiber_number}) is applicable to treat filament
  networks growing behind a skewed linear obstacle. (a) {\it Lab
    frame}: A generic skewed linear obstacle with skew angle
  $\varphi$. (b) {\it Obstacle frame}: The coordinate system is
  rotated by its skew angle $\varphi$ counter-clockwise. In this
  frame the initial skew of the system is expressed by an
  additional finite lateral motion at velocity $\tilde{v}_{\rm nw}^=$
  of an otherwise flat obstacle moving at $\tilde{v}_{\rm
    nw}^\perp$. (c) Parabolic (blue) and cosine (red) nonlinear
  obstacle shapes. The solid lines indicate the nonlinear obstacle
  shape, while the dashed line corresponds to a piecewise-linear
  triangular approximation. Horizontally the parabolic shape has a
  finite width, while the cosine obstacle shape is analyzed in
  periodic conditions laterally. (d-f) Piecewise-linear approximations
  of a parabola-like obstacle geometry. (d) The simplest approximation
  is a flat linear obstacle. (e) Due to the left-right symmetry in
  obstacle shape, in higher order approximations of the nonlinear
  obstacle (e.g. the triangular approximation shown here), only half
  of the number of sections have to be considered explicitly. The
  boundary condition at the center of the box has to be adjusted
  accordingly. (f) Subdividing the piecewise-linear sections again and
  again leads to ever higher accuracy in the approximation of the
  nonlinear obstacle shape and to ever smaller horizontal width of
  each subsection, $d_{{\rm br},i}^= = d_{\rm br}^= / n$. Approaching
  the continuum limit, $n \rightarrow \infty$, yields a PDE-model for
  nonlinear obstacle shapes.}
\label{figure4}
\end{figure}

It is not trivial to find an explicit expression for the outgrowth
rate out of the branching region behind a skewed linear obstacle,
which is rotated according to a constant skew angle $\varphi$, with
$-90^\circ < \varphi < +90^\circ$. However, for reasonably small skew
angle $\varphi$ and width of the orthogonal branching region $d_{\rm
  br}^\perp$ an approximation can be obtained within a rotated
coordinate frame, that is adjusted to the constant obstacle
skew. \fig{figure4}(a) shows a sketch of the obstacle geometry we are
interested in at this point. The branching region is given by a
rhomboid, with rotated (or skewed) upper and lower side. The
orthogonal width of the branching region $d_{\rm br}^\perp$ is again
defined parallel to the lateral sides of the obstacle, while its
horizontal width is given by $d_{\rm br}^=$. Network velocity $v_{\rm
  nw}$ is defined as before in the vertical direction. To find an
explicit expression for the rates of filament outgrowth, a rotation of
the coordinate frame to the point where the skew angle $\varphi$ of
the branching region vanishes simplifies the situation
(\fig{figure4}(b)). In this frame the obstacle appears flat
horizontally, very similar to the setup analyzed before in
\sec{nonperiodic_horizontal_obstacle_boundaries}. The finite skew angle $\varphi$, however,
manifests in a non-vanishing lateral network growth velocity
$\tilde{v}_{\rm nw}^=$, while the vertical propulsion speed of the
obstacle is the orthogonal part of the network velocity in this frame,
$\tilde{v}_{\rm nw}^\perp$. (Where applicable, we will denote
variables defined in the rotated coordinate frame by $\,\tilde{}\,$.)
In the following we will refer to the original coordinate system
\fig{figure4}(a) as the {\it lab frame} and to the rotated picture
\fig{figure4}(b) as the {\it obstacle frame}.

Using the transformation to the obstacle frame for a given skew angle
$\varphi$, we can now deduce the relevant parameters. The filament
orientation angles $\tilde{\theta}$ are given relative to the
obstacle, i.e. for $\tilde{\theta}=0^\circ$ filaments are growing
vertically in the rotated frame. The two components of obstacle growth
velocity are given by,
\begin{equation}
\tilde{v}_{\rm nw}^\perp(\varphi)=v_{\rm nw} \cos \varphi \textnormal{ and } \tilde{v}_{\rm nw}^=(\varphi) = v_{\rm nw} \sin \varphi\,.
\label{paricle_separation_vnw_box_si}
\end{equation}
The two obstacle diameters, which scale the vertical and horizontal outgrowth rates are,
\begin{equation}
\tilde{d}_{\rm br}^\perp(\varphi) = d_{\rm br}^\perp \cos \varphi \textnormal{ and } \tilde{d}_{\rm br}^=(\varphi) = \frac{d_{\rm br}^=}{\cos \varphi} \,.
\label{paricle_def_dbr_box_si}
\end{equation}
According to these definitions, we can now reformulate the continuum
description in the obstacle frame. The rate equation model
\eq{system_dgl_fiber_number} for the temporal evolution of orientation
dependent filament density remains unchanged with respect to the
transformed orientation angles $\tilde{\theta}$, however with adjusted
outgrowth rate $k_{\rm gr}(\theta,v_{\rm nw})$ given by the sum of
orthogonal outgrowth,
\begin{equation}
{\tilde{k}_{\rm gr}^{\perp}}(\tilde{\theta},\varphi,v_{\rm nw})=
\begin{cases} 
0 & \text{if $\vert\tilde{\theta}\vert \leq \tilde{\theta}_{\rm c}$}\\
\frac{\tilde{v}_{\rm nw}^\perp(\varphi)-v_{\rm fil} \cos{\tilde{\theta}}}{(\tilde{d}_{\rm br}^\perp(\varphi)/2)} &\text{if $\vert\tilde{\theta}\vert > \tilde{\theta}_{\rm c}$}
\end{cases}\,,
\label{particle_orth_outgr_boxsi}
\end{equation}
and the absolute value of lateral outgrowth,
\begin{equation}
{\tilde{k}_{\rm gr}^{=}}(\tilde{\theta},\varphi,v_{\rm nw})=
\begin{cases} 
\frac{\tilde{v}_{\rm nw}^=(\varphi) + \tilde{v}_{\rm nw}^\perp(\varphi) \tan{\tilde{\theta}}}{(\tilde{d}_{\rm br}^=(\varphi)/2)} & \text{if $\vert\tilde{\theta}\vert \leq \tilde{\theta}_{\rm c}$}\\
\frac{\tilde{v}_{\rm nw}^=(\varphi) + v_{\rm fil} \sin{\tilde{\theta}}}{(\tilde{d}_{\rm br}^=(\varphi)/2)} & \text{if $\vert\tilde{\theta}\vert > \tilde{\theta}_{\rm c}$}
\end{cases}\,,
\label{particle_lat_outgr_boxsi}
\end{equation}
with,
\begin{equation}
 \tilde{\theta}_{\rm c}=\arccos \left( \frac{\tilde{v}_{\rm nw}^\perp(\varphi)}{v_{\rm fil}} \right)\,.
\label{theta_crit_tilde}
\end{equation}
The lateral outgrowth rate compared to
\eq{particle_lat_outgrowth_rate} is biased by the finite horizontal
velocity of the network in the obstacle frame $\tilde{v}_{\rm
  nw}^=$. This additional feature also breaks the symmetry in the
resulting steady state filament orientation distributions. The sign of
the outgrowth rate determines, whether filaments are growing out to
the left ($\tilde{k}_{\rm gr}^{=}<0$) or the right ($\tilde{k}_{\rm
  gr}^{=}>0$) of the branching region, which is only relevant when
keeping track of the filaments' fate after outgrowth. This will become
important later in \sec{piecewise_linear_obstacle_approximation}, but
at this point the sign of lateral outgrowth is suppressed.

\begin{figure}[htp]
\centering
\includegraphics[width=0.9\textwidth]{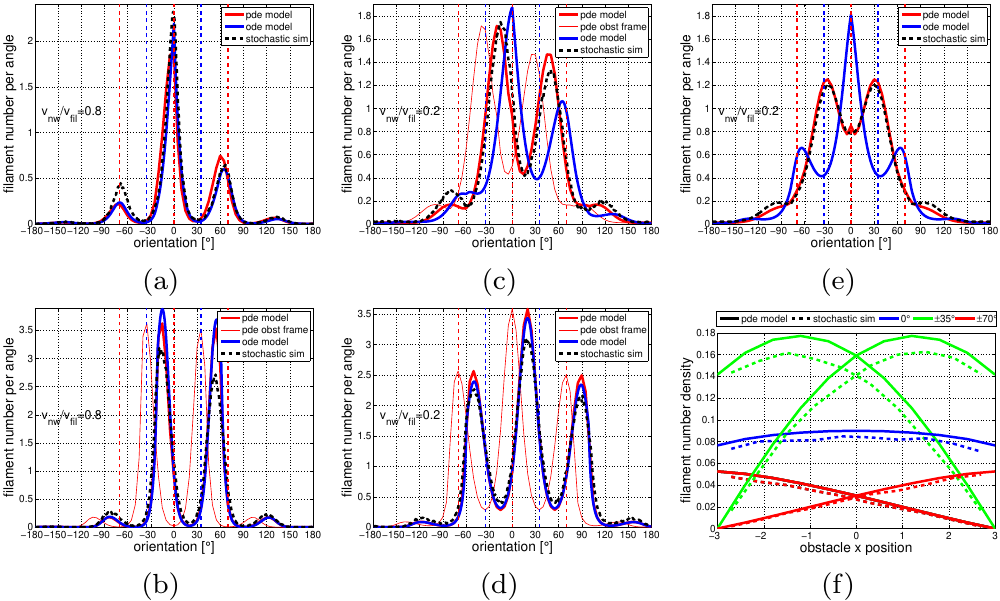}
\caption{(a-d) Representative results for the stationary filament
  orientation distributions of a linear tilted obstacle obtained in
  the PDE-model (thick red solid line), the ODE-model
  (\eq{system_dgl_fiber_number}) (blue solid line) and in stochastic
  network simulations (black dashed line). Where applicable also the
  transformation of the resulting PDE-model distributions to the
  obstacle frame is shown (thin red line). Here the orientation
  patterns can be interpreted as $\pm35$ and $+70/0/\!\!-\!\!70$
  patterns, that were also active in the tail of flat linear obstacles
  (cf. \fig{figure2} \& \fig{figure3}). The active parameters are,
  $\varphi=20^\circ$, $k_{\rm b}=20$, $d_{\rm br}^\perp=2\delta_{\rm
    fil}$ and (a) $r_{\rm br}=3$, $v_{\rm nw}/v_{\rm fil}=0.8$. (b)
  $r_{\rm br}=100$, $v_{\rm nw}/v_{\rm fil}=0.8$. (c) $r_{\rm br}=3$,
  $v_{\rm nw}/v_{\rm fil}=0.2$. (d) $r_{\rm br}=100$, $v_{\rm
    nw}/v_{\rm fil}=0.2$. (e) and (f) Comparison of the steady state
  solution for a nonperiodic linear flat obstacle from the ODE-model
  (\eq{system_dgl_fiber_number}), the PDE-model (\eq{pde_fiber_number}) and stochastic network simulations. (e) For small $r_{\rm br}=3$, the ODE-solution (blue line)
  predicts the absence of a $\pm35$ orientation distribution
  (cf. \fig{figure2}). However, for relatively low network velocity
  at $v_{\rm nw}/v_{\rm fil}=0.2$, the spatially resolved PDE-solution (red line)
  as well as network simulations (black
  dashed line) yield such a pattern. (f) The spatial
  filament distributions are far from uniform. Results from PDE-model
  (solid) and stochastic network simulation (dashed) rather show an
  accumulation of filaments at the lateral flanks of the obstacle, and
  therefore also horizontal outgrowth rates and as a direct
  consequence, the resulting orientation distribution averaged over
  the whole branching region differs from the prediction of the ODE-model.}
\label{figure5}
\end{figure}

To test our approximation, we again compared steady state filament
orientation distributions, obtained by numerical iteration of
\eq{system_dgl_fiber_number} within the obstacle frame and subsequent
transformation to the lab frame, against stochastic
simulations. Typical cases are shown in \fig{figure5}(a)--(d) for a
moderate obstacle skew angle of $\varphi=20^\circ$, fast and slow
network velocity in the lab frame $v_{\rm nw}$, and small and large
horizontal obstacle diameter as indicated by the length scale ratio
$r_{\rm br}$ (\fig{figure5} also shows results obtained with a PDE-model which we will introduce below). If the results are interpreted directly in the adjusted
obstacle frame, very similar orientation patterns as the familiar
$\pm35$ and $+70/0/\!\!-\!\!70$ peaked distributions are realized in
steady state. In the limit of very large length scale ratio $r_{\rm br}$, outgrowth
in the lateral direction can be neglected again and the resulting
patterns resemble orientation patterns for a network growing behind a
large obstacle, only now within the rotated obstacle frame
(cf. \fig{figure5}(b) \& (d)). For small $r_{\rm br}$ in combination
with a relatively high obstacle velocity, horizontal outgrowth cannot
be neglected and the orientation distribution has to be interpreted in
the lab frame (\fig{figure5}(a)). In between a transition occurs where
the solutions from stochastic simulations already have to be
interpreted in the obstacle frame, while results from the rate equation model
are still to be interpreted directly in the lab frame
(\fig{figure5}(c)). The reason for this is the spatial resolution in
the lateral filament position, which is introduced in filament based
stochastic simulations and neglected in the continuum model. Laterally nonuniform
spatial filament distributions lead to differences in the effective
outgrowth of filaments to the sides of the branching region. In order to include a similar 
spatial resolution in the continuum model, in the following we
will incorporate an additional advection term in the equation. In this
way it will also become possible to treat nonlinear obstacle shapes
within the continuum description.

\section{Nonlinear obstacle shape}
\label{nonlinear_obstacle_shape}

In this section, we focus on actin network dynamics in the tail of
curved (i.e. nonlinear) obstacle geometries. The definitions of the
relevant parameters stay the same in this geometry as introduced in
\fig{figure1}(a). \fig{figure1}(c) features a representative steady
state network from stochastic simulation.  In the following, we will
stepwise extend the continuum model \eq{system_dgl_fiber_number} with
the goal to also incorporate nonlinear obstacle shapes in the
description. We will do this first within an adjusted ordinary
differential equation (ODE) model in a piecewise-linear approximation
of the obstacle surface and then within a partial differential
equation (PDE) model, where an advection term will explicitly govern lateral
filament growth in the rate equation, additional to the reaction
terms.

\subsection{Piecewise-linear obstacle approximation}
\label{piecewise_linear_obstacle_approximation}

Nonlinear obstacle shapes can be treated within a
piecewise-linear approximation for the obstacle surface in which the
ODE-model from above is still applicable. The construction of this
approximation is sketched in \fig{figure4}(d)--(f) for a parabola-like
curved obstacle surface. The branching region is divided laterally in
$n$ sections of equal size $d_{{\rm br},i}^= =d_{\rm br}^= / n$
($i=1,\dots,n$). The orthogonal width of the branching region in each
section remains unchanged at $d_{{\rm br},i}^\perp = d_{\rm br}^\perp$
. The resulting obstacle approximation is a combination of lateral
sections with skewed linear shape according to skew angles
$\varphi_{i}$, very similar to the obstacles discussed before. The
individual sections (i.e. the ODE-model equation of each section) are
coupled to their direct neighbors via the lateral outgrowth rates
$k_{{\rm gr},i}^=$ to the left and right boundaries as indicated in
the sketch. In case of periodic lateral boundary conditions of the
obstacle, sections $i=1$ and $i=n$ are also coupled via
outgrowth. Individual outgrowth rates within the local obstacle frame
of each section can then be formulated according to
\eq{particle_orth_outgr_boxsi} and \eq{particle_lat_outgr_boxsi} as
before.

Assuming a single linear obstacle is the simplest approximation of a
nonlinear geometry (\fig{figure4}(d)). To increase accuracy,
the obstacle can be subdivided laterally in ever smaller
subsections. Due to the left-right symmetry of many biologically
relevant obstacle shapes (e.g. the parabola-like shape in the sketch)
it is sufficient to obtain results in one half-space of the obstacle
geometry. This translates into a problem of $n/2$ coupled ODEs for an
even number of subsections in the piecewise-linear approximation. In
order to exploit this symmetry, lateral outgrowth from the section
adjacent to the middle of the box has to be adjusted: Filaments that
would leave the section in the middle towards the other side of the
obstacle are reinjected again at this position with mirrored
orientation, $\theta \rightarrow - \theta$\,. For instance in case of
a triangular approximation of the nonlinear obstacle
(\fig{figure4}(e)), only the outgrowth behavior at one boundary needs
adjustment compared to the problem of a skewed linear obstacle shape
that was already discussed before. \fig{figure6}(a) and
\fig{figure7}(a), show steady state filament orientation distributions
for a laterally nonperiodic parabola and periodic-cosine shape
obtained by the ODE-model in triangular approximation
respectively. The nonlinear obstacle shapes together with their
respective triangular approximations are illustrated in
\fig{figure4}(c). To parametrize nonlinear parabola and cosine
geometries in this context, we are using the left hand side skew angle
of the triangular obstacle approximation $\bar{\varphi}$ in
combination with the horizontal width of the obstacle indicated by
$r_{\rm br}$.

\subsection{Reaction-advection equation}

The continuum limit of an infinitely large number of piecewise-linear
subsections, each of infinitesimal width, yields a PDE-model for
nonlinear obstacle shapes. The model equations in this limit are
complemented by an additional filament advection term, similar to
hydrodynamic models, that incorporates lateral filament growth in
\eq{system_dgl_fiber_number},
\begin{equation}
\begin{split}
\frac{\partial N(\theta,x,t)}{\partial t} + \frac{\partial \left(N(\theta,x,t) \cdot v_{\rm fil}^=(\theta,\varphi,v_{\rm nw}) \right) }{\partial x} = \hat{k}_{\rm b}\int \mathcal{W}(\theta,\theta') N(\theta',x,t) \,\mbox{d}\theta' \\
- k_{\rm c} N(\theta,x,t) - k_{\rm gr}^\perp(\theta,\varphi,v_{\rm nw}) N(\theta,x,t) \,,
\end{split}
\label{pde_fiber_number} 
\end{equation}
with the horizontal filament growth speed, $v_{\rm
  fil}^=(\theta,\varphi,v_{\rm nw})$, and the vertical outgrowth rate
in the lab frame, $k_{\rm gr}^\perp(\theta,\varphi,v_{\rm nw})$. Both,
filament growth velocity and outgrowth rate are functions of filament
orientation $\theta$ as well as the local obstacle skew angle
$\varphi(x)$, which is active at lateral position $x$. Here again, we
average outgrowth of filaments in the vertical direction by an
effective rate, while horizontal growth is now spatially resolved due
to the additional advection term.

For arbitrary obstacle shape, that can be expressed in terms of an analytic
function $o(x)$, the local skew angle can be written as,
\begin{equation}
\varphi(x) = -\arctan \left(\frac{{\partial} o(x)}{{\partial} x} \right)\,.
\label{loc_obst_skew}
\end{equation}
The two components of the filament growth velocities within the {\it
  local} obstacle frame can thus be written as,
\begin{equation}
\tilde{v}_{\rm fil}^\perp (\tilde{\theta},\varphi,v_{\rm nw}) =
\begin{cases} 
0 & \text{if $\vert\tilde{\theta}\vert \leq \tilde{\theta}_{\rm c}$}\\
\tilde{v}_{\rm nw}^\perp(\varphi)-v_{\rm fil} \cos{\tilde{\theta}} &\text{if $\vert\tilde{\theta}\vert > \tilde{\theta}_{\rm c}$}
\end{cases}\,,
\label{growth_vel_vert}
\end{equation}
and,
\begin{equation}
\tilde{v}_{\rm fil}^= (\tilde{\theta},\varphi,v_{\rm nw}) =
\begin{cases} 
\tilde{v}_{\rm nw}^=(\varphi)+\tilde{v}_{\rm nw}^\perp(\varphi) \tan{\tilde{\theta}} & \text{if $\vert\tilde{\theta}\vert \leq \tilde{\theta}_{\rm c}$}\\
\tilde{v}_{\rm nw}^=(\varphi)+v_{\rm fil} \sin{\tilde{\theta}} & \text{if $\vert\tilde{\theta}\vert > \tilde{\theta}_{\rm c}$}
\end{cases}\,,
\label{growth_vel_par}
\end{equation}
with the components of the obstacle velocity, $\tilde{v}_{\rm
  nw}^\perp(\varphi)$ and $\tilde{v}_{\rm nw}^=(\varphi)$, defined as
in \eq{paricle_separation_vnw_box_si} and $\tilde{\theta}_{\rm c}$ as
in \eq{theta_crit_tilde}. A transformation (i.e. rotation) from the
local obstacle frame into the local lab frame at each horizontal
position along the obstacle surface subsequently leads to filament
growth velocities, which enter the PDE-model \eq{pde_fiber_number} due
to vertical outgrowth and advection,
\begin{equation}
\begin{array}{rccl}
v_{\rm fil}^\perp (\theta,\varphi,v_{\rm nw}) & = & &   \tilde{v}_{\rm fil}^\perp (\tilde{\theta},\varphi,v_{\rm nw}) \cos \varphi + \tilde{v}_{\rm fil}^= (\tilde{\theta},\varphi,v_{\rm nw}) \sin \varphi \\
v_{\rm fil}^= (\theta,\varphi,v_{\rm nw})     & = &-&  \tilde{v}_{\rm fil}^\perp (\tilde{\theta},\varphi,v_{\rm nw}) \sin \varphi + \tilde{v}_{\rm fil}^= (\tilde{\theta},\varphi,v_{\rm nw}) \cos \varphi \,.
\end{array}
\label{trafo_vel_lab_frame}
\end{equation}
Note, that filaments with a positive $v_{\rm fil}^\perp
(\theta,\varphi)$ are growing vertically in opposite direction to the
obstacle.  Using these expressions, we can now rewrite the advection
term in \eq{pde_fiber_number} as,
\begin{equation}
\frac{\partial \left(N \cdot v_{\rm fil}^= \right) }{\partial x} = v_{\rm fil}^= \frac{\partial N}{\partial x} + N \frac{\partial  v_{\rm fil}^=}{\partial x} = v_{\rm fil}^= \frac{\partial N}{\partial x} + N \frac{\partial  v_{\rm fil}^=}{\partial \varphi}\frac{\partial \varphi}{\partial x}\,,
\label{advection_term}
\end{equation}
with,
\begin{equation}
\frac{\partial  v_{\rm fil}^=}{\partial \varphi} = - \frac{\partial \left( \tilde{v}_{\rm fil}^\perp \sin \varphi \right)}{\partial \varphi} + \frac{\partial \left( \tilde{v}_{\rm fil}^= \cos \varphi \right)}{\partial \varphi} \,,
\label{partial_v_lat}
\end{equation}
and substituting \eq{growth_vel_vert},
\begin{equation}
\frac{\partial \left( \tilde{v}_{\rm fil}^\perp \sin \varphi \right)}{\partial \varphi} =
\begin{cases} 
0 & \text{if $\vert\tilde{\theta}\vert \leq \tilde{\theta}_{\rm c}$}\\
- v_{\rm nw} \sin^2 \varphi + v_{\rm nw} \cos^2 \varphi - v_{\rm fil} \cos \tilde{\theta} \cos \varphi - v_{\rm fil} \sin \varphi \sin \tilde{\theta} & \text{if $\vert\tilde{\theta}\vert > \tilde{\theta}_{\rm c}$}
\end{cases}\,,
\label{partial_v_orth_sin}
\end{equation}
and \eq{growth_vel_par},
\begin{equation}
\begin{array}{ll}
\frac{\partial \left( \tilde{v}_{\rm fil}^= \cos \varphi \right)}{\partial \varphi} = & \\
\begin{cases} 
v_{\rm nw} \cos^2 \varphi - v_{\rm nw} \sin^2 \varphi - 2 v_{\rm nw} \cos \varphi \sin \varphi \tan \tilde{\theta} -v_{\rm nw} \cos^2 \varphi \cos^{-2} \tilde{\theta} & \text{if $\vert\tilde{\theta}\vert \leq \tilde{\theta}_{\rm c}$}\\
v_{\rm nw} \cos^2 \varphi -v_{\rm nw} \sin^2 \varphi - v_{\rm fil} \sin \tilde{\theta} \sin \varphi - v_{\rm fil} \cos \varphi \cos \tilde{\theta} & \text{if $\vert\tilde{\theta}\vert > \tilde{\theta}_{\rm c}$}
\end{cases}\,,
\end{array}
\label{partial_v_lat_cos}
\end{equation}
and using \eq{loc_obst_skew},
\begin{equation}
\frac{\partial \varphi}{\partial x} = \frac{- \frac{{\partial^2} o(x)}{{\partial} x^2}}{1 + \left(\frac{{\partial} o(x)}{{\partial} x}\right)^2}\,.
\label{partial_phi}
\end{equation}
As we average outgrowth in the vertical direction by the rate $k_{\rm
  gr}^\perp(\theta,\varphi,v_{\rm nw})$ in the lab frame, while
horizontal growth is now spatially resolved, we have to take into
account the finite lateral movement of filaments along the obstacle
geometry, when considering filaments growing out of the branching
region vertically. To approximate the filament outgrowth rate
accordingly, we again assume a locally linear obstacle shape at each
horizontal position to find an analytical expression for the outgrowth
rate in \eq{pde_fiber_number},
\begin{equation}
k_{\rm gr}^\perp(\theta,\varphi,v_{\rm nw}) = \sqrt{ \frac{{v_{\rm fil}^\perp}^2 + {v_{\rm fil}^=}^2}{\left(\frac{d_{\rm br}^\perp/2}{v_{\rm fil}^\perp/v_{\rm fil}^= - \tan \varphi} \right)^2+\left(\frac{d_{\rm br}^\perp/2}{1 - v_{\rm fil}^=/v_{\rm fil}^\perp \tan \varphi}\right)^2} } \,.
\label{outgrowth_pde}
\end{equation}

To calculate steady state filament orientation distributions from this
PDE-model, \eq{pde_fiber_number}--\eq{outgrowth_pde} can be numerically
iterated in discretized space and time using for instance a second
order upwind scheme for the spatial differential in
combination with an Euler method for temporal iteration
\cite{press1992}. For this procedure, we are using a uniform filament
distribution in space and orientation as initial condition and,
dependent on the specific obstacle shape considered, either periodic
boundary conditions or zero filament density at the respective inflow
boundary.

\subsection{Linear obstacle shape revisited}

The important advantage of using the PDE-model \eq{pde_fiber_number}
compared to the initial ODE \eq{system_dgl_fiber_number} to solve for
steady state filament patterns clearly lies in its additional spatial
resolution. This benefit not only increases the applicability of the
model to a broader range of obstacle shapes, but also manifests itself when
solving for specific parameter combinations for simple nonperiodic
linear flat obstacles that have been already accessible using the
initial ODE equations. \fig{figure5}(e) illustrates the steady state
filament orientation distributions averaged over the whole branching
region behind an obstacle of small width (i.e. $r_{\rm br}=3$). For
such laterally small and flat obstacles, the ODE-model predicted the
absence of the $\pm35$ orientation pattern for arbitrary obstacle velocities
$v_{\rm nw}$ (cf. \fig{figure2}). Due to their additional spatial
resolution horizontally, the PDE-model as well as results from
stochastic network simulations yield this orientation distribution
nevertheless for a small range of parameters at relatively slow
obstacle velocity, $v_{\rm nw}/v_{\rm fil}=0.2$. As shown in
\fig{figure5}(f), the spatial steady state filament distributions from
the PDE-model are nonuniform horizontally and in good agreement
to results from stochastic simulations. This has an impact on the
effective lateral outgrowth of filaments in the model and thus is able
to change the final (spatially averaged) filament distributions
characteristically as shown in this example.

\subsection{Parabolic and periodic-cosine obstacle shapes}
\label{parabolic_and_periodic_cosine_obstacle_shapes}

The PDE-model allows us to systematically analyze actin networks in
the tail of nonlinear obstacle shapes. As a proof of principle, in the
following we have chosen two different obstacle shapes which are
motivated by highly relevant biological examples. On the one hand, we
model a laterally nonperiodic parabolic obstacle shape (blue solid
line in \fig{figure4}(c)) which is a first approximation for the
spherical or ellipsoidal shape of actin-propelled intracellular
pathogens. On the other hand, actin growth behind a periodic-cosine
geometry (red solid line in \fig{figure4}(c)) is analyzed, motivated
by the laterally widely spread but corrugated leading edge of a
crawling or spreading cell.

\begin{figure}[htp]
\centering
\includegraphics[width=0.9\textwidth]{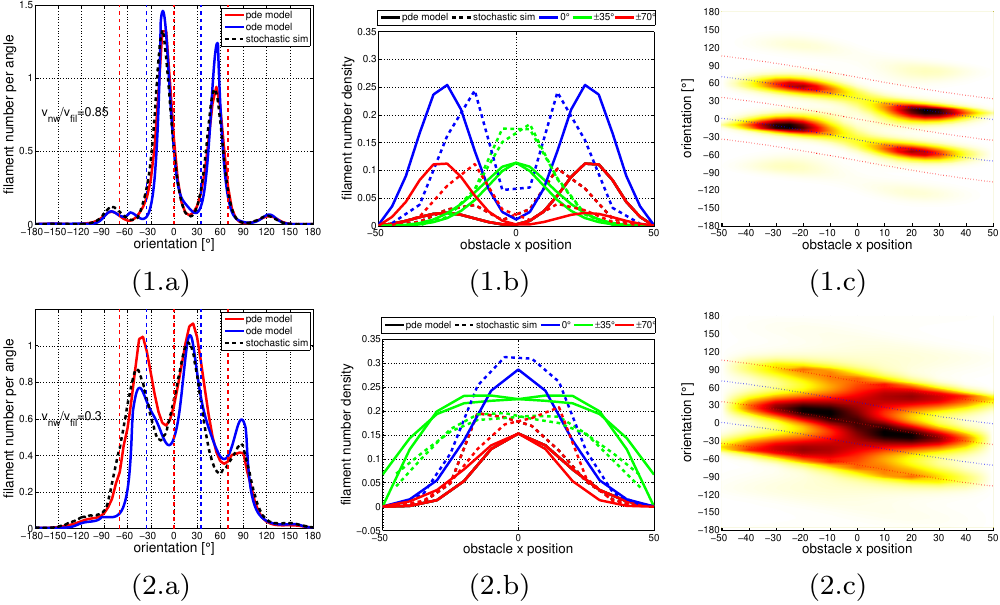}
\caption{Typical stationary filament distributions in the tail of a
  nonperiodic parabolic obstacle (cf. blue solid line in
  \fig{figure4}(c)) for the parameter combinations,
  $\bar{\varphi}=20^\circ$, $r_{\rm br}=50$, $d_{\rm
    br}^\perp=2\delta_{\rm fil}$, $k_{\rm b}=20$, $k_{\rm c}=0.05$;
  (1) $v_{\rm nw}/v_{\rm fil}=0.85$; (2) $v_{\rm nw}/v_{\rm
    fil}=0.3$. (a) Filament orientation distributions spatially
  averaged laterally over the left hand side half space behind the
  symmetric obstacle. For comparison the results from PDE-model,
  ODE-model in triangular approximation (as illustrated by the dashed blue
  line in \fig{figure4}(c)) and stochastic simulations are shown. (b)
  Spatial distributions of characteristic filament angles at
  $\theta=0^\circ$, $\pm35^\circ$ and $\pm70^\circ$. Interpretation in
  the lab frame indicates coexistence of the competing $\pm35$ and
  $+70/0/\!\!-\!\!70$ patterns at different lateral positions along
  the obstacle. (c) Heat map of the spatially resolved filament
  orientation distributions, where cooler colors indicate increasing
  filament density. Plotting dashed lines at characteristic angles at
  $\tilde{\theta}=0^\circ$, $\pm35^\circ$ and $\pm70^\circ$ in the
  obstacle frame indicates that in this frame a single familiar pattern 
  dominates. The apparent coexistence of different patterns
  is artificially introduced due to a misinterpretation of the results
  in the lab frame.}
\label{figure6}
\end{figure}

\begin{figure}[htp]
\centering
\includegraphics[width=0.9\textwidth]{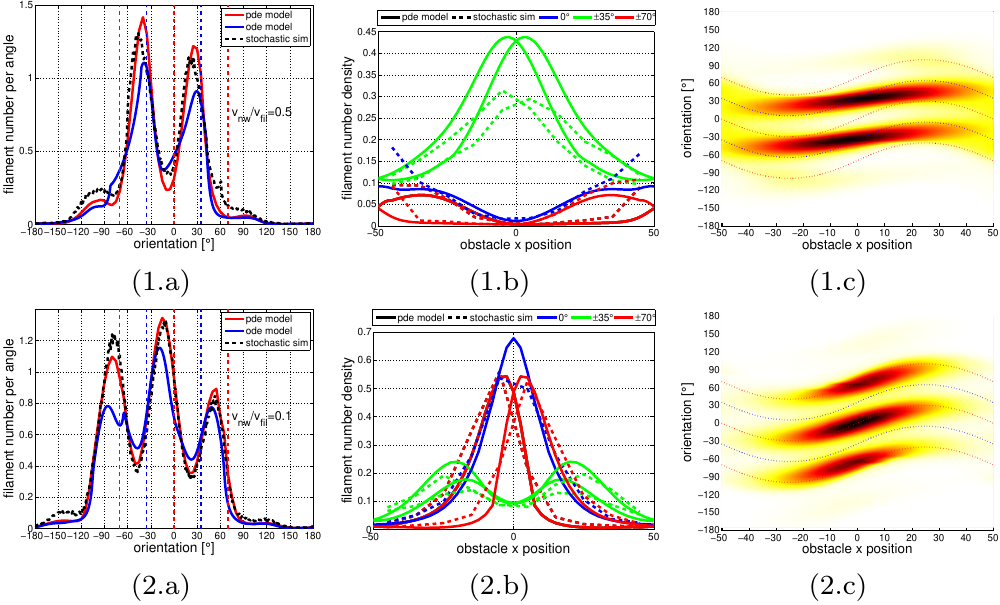}
\caption{Same as \fig{figure6} for a periodic-cosine shaped obstacle
  as illustrated by the solid red line in \fig{figure4}(c). Active
  parameters are $\bar{\varphi}=-20^\circ$, $r_{\rm br}=50$, $d_{\rm
    br}^\perp=2\delta_{\rm fil}$, $k_{\rm b}=20$, $k_{\rm c}=0.05$;
  (1) $v_{\rm nw}/v_{\rm fil}=0.5$; (2) $v_{\rm nw}/v_{\rm
    fil}=0.1$. Again an apparent coexistence of competing patterns disappears when results are interpreted in the obstacle
  frame, that is locally orthogonal to the obstacle surface.}
\label{figure7}
\end{figure}

In \fig{figure6} and \fig{figure7}, the resulting steady state
patterns from the three different modeling approaches (PDE-model,
ODE-model in triangular piecewise-linear approximation and stochastic
network simulations) are shown for specific parameter combinations. In
(a) the filament orientation distributions spatially integrated over
the left side of the symmetric obstacle are shown. Here, the PDE-model
corresponds very well to stochastic network simulations. Despite the
rough piecewise-linear approximation of the nonlinear obstacle shape
in the ODE-model, the results capture the overall characteristic of
the resulting network pattern very well. Subfigures (b) illustrate the
resulting spatial distributions of certain characteristic filament
angles. For specific parameter cases separate spatial domains emerge
over the horizontal obstacle dimension, in which the signature of the
two alternative steady state network patterns,
i.e. $+70/0/\!\!-\!\!70$ and $\pm35$, are observed. For a better
overview over the complete spatially resolved orientation patterns
from the PDE-model, \fig{figure6}(c) and \fig{figure7}(c) illustrate
the two dimensional filament distribution in heat plots, where regions
of cooler color indicate higher filament density. The apparent
coexistence of alternative patterns resolves when additional dotted
lines are included in the heat map to indicate local filament
orientations that correspond to the characteristic orientations
$\tilde{\theta}=0^\circ,\,\pm35^\circ,\,\pm70^\circ$ within the
obstacle frame, locally orthogonal to the obstacle surface along its
contour. Although coexistence of alternative filament orientations is
present in the lab frame, an interpretation in the local obstacle
frame yields one unique pattern only that is present along the lateral
extension of the obstacle.

\section{Discussion}

In this article, we have introduced several theoretical approaches at
different levels of complexity with the common goal to predict actin
filament orientation patterns at the surface of stiff two dimensional
obstacles, whose surfaces promote growth of actin networks. This
problem is not only central for many important health and disease related biological phenomena,
such as migration of animal cells or propagation
of pathogens in their host, it is also a prominent example for a
physical process whose properties emerge on a mesoscopic length scale
in relative independence of the details of the underlying molecular
processes. For example, the competition of the two fundamentally
different network architectures discussed here does not rely on the
exact value of the branching angle, but rather on the fact
that any branching angle below 90 degrees can lead to the possibility
that fundamentally different network architectures satisfy the simultaneous requirements of forward growth and side branching. Our model focuses on the geometrical aspects of this situation (in particular filament orientation and obstacle shape). In the future, it might be extended by other important aspects of this complex biological system, including the details of the filament-membrane interaction and the role of filament bending \cite{zimmermann2012,zhu2012}.

In general, all our results were benchmarked against stochastic
computer simulations which in principle can be used to include many
details of the underlying molecular processes.  Here we have adopted
the established view that polymerization, branching and capping are
the dominating processes in the context of growing actin gels.  For
relatively simple linear obstacle shapes, a rate equation (ODE) model
yielded accurate results and, due to its reduced complexity, also
allowed for analytical progress. Using this approach, we found that
either of two competing orientation patterns emerges in steady state
and transitions between the two are triggered mostly by changes in the
velocity of the obstacle. A similar result has already been obtained
in earlier studies for a flat obstacle with large lateral extension,
where filament orientation distributions with characteristic peaks at
either $+70/0/\!\!-\!\!70$ or $\pm35$ have been found to dominate the
steady state of growing networks \cite{maly2001,weichsel2010_2}. These
results were now extended to finite sized obstacles. We find that for
very small objects, the $+70/0/\!\!-\!\!70$ pattern is dominant, while
for larger objects, mutual stability of the two
patterns recovers. In case of obstacles with a tilted straight contour (linear
obstacles), again very similar patterns are predicted. However, for
most parameter configurations they have to be interpreted within a
rotated obstacle frame.

Curved (nonlinear) obstacle shapes have been analyzed in the
ODE-model using a piecewise-linear approximation of the given
geometry. As an alternative, a continuous spatial coordinate in the
horizontal direction was introduced to the model equation by
incorporating a filament advection term, thus yielding a
PDE-model. The additional benefit of this spatial resolution clearly
lies in the gain of accuracy compared to results obtained earlier. For
specific parameter combinations, it was shown that the lack of spatial
resolution can lead to incorrect predictions regarding the resulting
filament patterns in steady state. Analysis of nonlinear obstacle shapes in the PDE-model yielded an
apparent coexistence of the two competing network patterns at
different horizontal positions along the obstacle surface in the lab
frame. A transformation to the obstacle frame, locally at each lateral
position, however revealed that one unique orientation pattern is
dominating the network structure as before. Again, these results
are in good agreement with computer simulations.

In order to experimentally test these theoretical predictions, the
method of choice would be electron tomography
\cite{urban2010,weichsel2012,vinzenz2012}, although in the future
super-resolution microscopy like dual-objective STORM might become an
interesting alternative \cite{xuk2012}. In the context of a
rapidly increasing image quality of electron microscopy (EM) data for
actin networks, the correct analysis and interpretation of the
measured observables becomes increasingly important. Therefore, a
detailed understanding of the structural network characteristics
emerging under different situations is indispensable. In this work, we
have also shown that the extracted orientation distribution of actin
filament networks from experimental EM images growing at the surface of
nonuniform obstacles can be easily misinterpreted. We have shown that
a standard measurement in the lab frame would yield apparently novel
patterns that could not be explained by existing models. For a correct
interpretation of such findings, we introduced a rotated obstacle
frame which lies locally orthogonal to the obstacle surface. Within
this reference frame, the seemingly novel filament orientation
patterns are rationalized and can be understood in terms of the two
well known filament distributions, peaked at $+70/0/\!\!-\!\!70$ or
$\pm35$. Thus our results also contribute to improving the way
experimental data can be compared to theoretical predictions.

A convenient setup to test our predictions would be the combination of
electron tomography with biomimetic assays with particles of various
shapes, which earlier have been analyzed mainly in regard to
macroscopic variables such as propulsion velocity and shape of the
comet tail \cite{schwartz2004}. Apart from obstacle shape, the
obstacle velocity relative to the filament polymerization speed is a central parameter in the model
which triggers transitions between filament orientation patterns. This
velocity could be adjusted for instance by applying a force against
network growth diminishing steady state growth. Changing the
biochemical reaction rates for capping and branching is expected to be of minor importance and
could be tested by varying the concentration of purified protein
solutions in the assay. As an additional benefit of introducing
different obstacle geometries in experiment, this might also lead to a
deeper understanding of Arp2/3 activation close to the surface of the
obstacle, which is one of the most important questions still to be
clarified in the context of actin-driven motility.

\begin{acknowledgments}
This work was supported by the BMBF FORSYS project Viroquant.  JW is
supported by the Deutsche Forschungsgemeinschaft (DFG grant We
5004/2-1). USS is member of the cluster of excellence CellNetworks at
Heidelberg University.
\end{acknowledgments}





\newpage

\end{document}